\documentclass[aps,prd,preprint,showpacs,nofootinbib,groupedaddress]{revtex4-1}
\usepackage{amsmath}
\usepackage{latexsym}
\usepackage{amsfonts}
\usepackage{graphicx}
\usepackage{hyperref}

\begin{document}

\title{The tension on the cosmological parameters from different observational data}

\author{Qing Gao}
\email{gaoqing01good@163.com}
\affiliation{MOE Key Laboratory of Fundamental Quantities Measurement, School of Physics, Huazhong University of Science and Technology, Wuhan 430074, China}
\author{Yungui Gong}
\email{yggong@mail.hust.edu.cn}
\affiliation{MOE Key Laboratory of Fundamental Quantities Measurement, School of Physics, Huazhong University of Science and Technology, Wuhan 430074, China}
\affiliation{Institute of Theoretical Physics, Chinese Academy of Sciences, Beijing 100190, China}

\begin{abstract}
Planck measurements of the cosmic microwave background power spectra find a lower value of the Hubble constant $H_0$
and a higher value of the fractional matter energy density $\Omega_{m0}$ for the concordance $\Lambda$CDM model, and
these results are in tension with other measurements.
The {\em Planck} group argued that the tension
came either from some sources of unknown systematic errors in some astrophysical measurements
or the wrong $\Lambda$CDM model applied in fitting the data.
We studied the reason for the tension on $H_0$ from different measurements
by considering two dynamical dark energy models. We found that
there is no tension between different data,
the constraint on $H_0$ is almost unchanged for different
dark energy models and the tension with the local measurements remains when the error bar on $H_0$ is tightened to be around 1.
We argue that the tension on $H_0$ is not caused by the fitting model.
\end{abstract}

\preprint{arXiv: 1308.5627}
\maketitle

\section{Introduction}

Ever since the discovery of the cosmic acceleration found by the observations
of type Ia supernovae (SNe Ia) in 1998 \cite{acc0,hzsst98,scpsn98},
an exotic energy component dubbed as dark energy with negative pressure which contributes
about 72\% to the total energy density in the universe has been proposed. Alternatively,
modified theories of gravity such as the Dvali-Gabadadze-Porrati model \cite{dgp},
f(R) gravity \cite{Carroll:2003wy,Starobinsky:2007hu,Hu:2007nk},
dRGT ghost-free massive gravity \cite{massgrav,Gong:2012yv},
are also used to explain the cosmic acceleration.
Although the cosmological constant is the
simplest candidate for dark energy and is consistent with current
observations, dynamical dark energy models such as the quintessence model
\cite{peebles88,wetterich88,quintessence,track1,track2} are also explored due to the many
orders of magnitude discrepancy between the theoretical estimation
and astronomical observations for the cosmological constant.
For a recent review of dark energy, please see Ref. \cite{sahni00,copelandde,Padmanabhan:2007xy,limiaode}.

Since the nature of dark energy is still unknown and a self-consistent dark energy theory is still
unavailable, current observational data provides us the only possible way to probe the nature of dark energy.
A lot of analyses by using the observational data have been done in the literature
\cite{alam04a,alam04b,barger,clarkson,corasaniti,astier,efstathiou,gerke,cooray,flux1,weller01,%
huang,star,lampeitl,corray9,gong10a,gong10b,gongcqg10,gong11,gongmnras13,li11, yu11,cai11,wetterich04,cpl1,cpl2,jbp05,zhu,Farooq:2012ev,Farooq:2013hq,Zhang:2013hma,Yahya:2013xma,Benitez:2013tfa}.
In particular, one usually parameterizes the equation of state parameter $w(z)$ of dark energy
with several parameters, such as the Chevallier-Polarski-Linder (CPL) parametrization
$w(a)=w_0+w_a(1-a)$ with two parameters $w_0$ and $w_a$ \cite{cpl1,cpl2}
and the SSLCPL model with explicit degeneracy relation between $w_0$ and $w_a$ \cite{gong1212,Gong:2013bn}.
Because of the degeneracies among the parameters $\Omega_{m0}$, $w_0$ and $w_a$ in the model,
complementary cosmological observations are needed to break the
degeneracies. The measurements on the cosmic microwave background
anisotropy, the baryon acoustic oscillation (BAO) measurements and the SNe Ia observations provide
complementary data.

Recently, {\em Planck} released its first year results which give
a higher value of fractional matter energy density $\Omega_{m0}=0.315\pm 0.017$
and a lower value of the Hubble constant $H_0=67.3\pm 1.2$ km$\,$s$^{-1}$$\,$Mpc$^{-1}$~\footnote{We refer this value as the {\em Planck} result}
for the concordance $\Lambda$CDM model \cite{planck13,Ade:2013zuv}.
However, Tammannk {\em et al} gave a lower value
$H_0=62.3\pm 1.3$(random)$\pm 4$(systematic) km$\,$s$^{-1}$$\,$Mpc$^{-1}$ by using the Cepheid-calibrated luminosity
of SNe Ia \cite{Tammann:2008xf} which differs at about $2.4\sigma$ from the
result $H_0=73.8\pm 2.4$ km$\,$s$^{-1}$$\,$Mpc$^{-1}$\footnote{This value is referred as the local measurements}
by calibrating the magnitude-redshift relation
of 253 SNe Ia with over 600 Cepheid variables \cite{Riess:2011yx} due to the difference of the calibration
of zero points. Applying the revised geometric maser distance to NGC 4258 to the
same Cepheid data used in \cite{Riess:2011yx},
the Hubble constant was lowered to be $H_0=70.6\pm 3.3$ km$\,$s$^{-1}$$\,$Mpc$^{-1}$ \cite{Efstathiou:2013via}.
If strong metallicity prior was imposed and three different distance anchors were combined,
the Hubble constant was measured as $H_0=72.5\pm 2.5$ km$\,$s$^{-1}$$\,$Mpc$^{-1}$ \cite{Efstathiou:2013via}.
By measuring the time delays between multiple images of lensed sources with free-form
modelling of gravitational lenses, the Hubble constant was
determined to be $H_0=69\pm 6$(statistic)$\pm 4$(systematic) km$\,$s$^{-1}$$\,$Mpc$^{-1}$ \cite{Sereno:2013ona}.
It seems that the calibration of zero point of SNe Ia data is still a controversial issue.
The {\em Planck} result disagrees with the local measurements at about $2.5\sigma$ level.
The {\em Planck} group argued that the tension
came either from some sources of unknown systematic errors in some astrophysical measurements
or the wrong $\Lambda$CDM model applied in fitting the data \cite{Ade:2013zuv}.
Despite of the tension on $H_0$ for the concordance $\Lambda$CDM model between the {\em Planck} constraint
and the local distance ladder measurements with SNe Ia data,
the {\em Planck} result is consistent with the results from BAO data
\cite{6dfgs,wjp,Anderson:2012sa,wigglez,ngbusca,Addison:2013haa} and the Hubble parameter $H(z)$
data \cite{hz1,hz2,hz1a,hz3}.
This excludes the possibility that the tension is due to the different redshift regions covered by different data.
In \cite{Spergel:2013rxa}, the authors found that some of the tensions come from
the 217 GHz$\times$217 GHz detector set spectrum, and they found $\Omega_{m0}=0.302\pm 0.015$
and $H_0=68.1\pm 1.1$ km$\,$s$^{-1}$$\,$Mpc$^{-1}$ by using a map-based foreground cleaning procedure.
By considering the effect of the cosmic variance on the measurement of the local Hubble constant \cite{Marra:2013rba}
or additional sterile neutrinos \cite{Wyman:2013lza},
it was shown that the tension can be partially relieved.

For the concordance $\Lambda$CDM model, the best fitting value from
the three year Supernova Legacy Survey (SNLS3) SNe Ia data \cite{snls3} is $\Omega_{m0}=0.227_{-0.035}^{+0.042}$ \cite{Ade:2013zuv},
and the constraint by the Union2.1 SNe Ia data is $\Omega_{m0}=0.295^{+0.043}_{-0.040}$ \cite{union2.1}.
With the improved photometric calibration of
740 SNe Ia (JLA data) obtained from the joint analysis of the Sloan Digital Sky Survey (SDSS)-II and SNLS collaborations \cite{jla},
the constraint becomes $\Omega_{m0}=0.295\pm 0.034$ which is consistent
with both Union2.1 and the {\em Planck} results. It seems that the
tension on $\Omega_{m0}$ between the SNLS3 SNe Ia constraint
and the {\em Planck} constraint (at about $2\sigma$ level) is due
to the systematics of the calibration of the SNLS3 SNe Ia data, so no such tension from different measurements exists.

However, for different data it is still worthy studying the dependence of
the parameter constraints on dynamical dark energy models and their consistencies with $\Lambda$CDM model.
For the dark energy model with constant equation of state ($w$CDM model),
{\em Planck} data alone gave $\Omega_{m0}=0.204^{+0.052}_{-0.051}$, $w=-1.49^{+0.30}_{-0.29}$
and $H_0=85.0\pm 10.9$ \cite{Addison:2013haa}
which is in tension with the flat $\Lambda$CDM model with $w=-1$ at more than $1\sigma$ level.
If the $H_0$ prior is combined with {\em Planck} data, $w=-1.24^{+0.18}_{-0.19}$ at 95\% confidence level
which is in tension with the flat $\Lambda$CDM model at more than $2\sigma$ level \cite{Ade:2013zuv}.
Fitting the CPL model to the combined BAO and {\em Planck} data, it was found that
$w_0=-1.04^{+0.72}_{-0.69}$ and $w_a<1.32$ at 95\% confidence level, and $\Lambda$CDM model
with $w_0=-1$ and $w_a=0$ is consistent with both the {\em Planck}+WP+BAO
\footnote{The meaning of different data combination is explained in the next section}
and the {\em Planck}+WP+Union2.1 combination but is in tension
with the {\em Planck}+WP+SNLS combination at about the $2\sigma$ level \cite{Ade:2013zuv}.
By using a different method of dark energy perturbation, it was found that $\Lambda$CDM model is consistent
with the data combination of {\em Planck}+WP+SNLS+BAO \cite{Xia:2013dea}.
In \cite{Cheng:2013csa}, the authors considered the CPL model with the addition
of dark radiation and the imposing of the $H_0=73.8\pm 2.4$ prior and they found that the concordance $\Lambda$CDM model is disfavoured
at more than $1\sigma$ level. Since the discrepancy between the {\em Planck} result and the local measurements is at about the $2.5\sigma$
level, the use of $H_0$ prior is inconsistent and the flat prior ([-2, 2])
on $w_a$ used in \cite{Ade:2013zuv,Cheng:2013csa} may be too restrictive.
We would like to address the issue whether the tension can be relieved
by using the dynamical dark energy models.

In this paper, we first apply a one parameter SSLCPL model
with explicit relation between $w_0$ and $w_a$ which models a wide class of thawing scalar field
over a large redshift region, then we use the CPL parametrization with
the 8 BAO data \cite{6dfgs,wjp,Anderson:2012sa,wigglez,ngbusca} and the 21 Hubble parameter $H(z)$ data \cite{hz1,hz2,hz1a,hz3}
in combination of three different SNe Ia samples: the SNLS3 sample of 472 SNe Ia data \cite{snls3},
the Union2.1 sample of 580 SNe Ia data \cite{union2.1}, and the JLA sample of 780 SNe Ia data from the joint analysis of SNLS
and SDSS-II collaborations \cite{jla}, and the
cosmic microwave background anisotropy data from the combination of {\em Planck}
temperature power spectrum with the WMAP polarization low-multipole likelihood data \citep{planck13,Ade:2013zuv,Bennett:2012zja}.

\section{observational data}

The SNLS3 SNe Ia data consist of 123 low-redshift SNe Ia with $z\lesssim 0.1$
mainly from Calan/Tololo, the Harvard-Smithsonian Center for Astrophysics (CfA) survey releases CfAI, CfAII and CfAIII,
and Carnegie Supernova Project (CSP),
242 SNe Ia over the redshift range $0.08<z<1.06$ observed from the SNLS \cite{snls3},
93 intermediate-redshift SNe Ia with $0.06\lesssim z\lesssim 0.4$ observed during the first season of SDSS-II supernova survey \cite{sdss2}, and 14 high-redshift SNe Ia with $z\gtrsim 0.8$
from the Hubble Space Telescope (HST) \cite{riessgold}.
The SNLS3 SNe Ia data used the combination of SALT2 and SiFTO light-curve fitters \cite{snls3}.
We also include the correction on the dependence of the host-galaxy stellar mass.
For the fitting to the SNLS3 data, we need to add two more nuisance parameters $\alpha$ and $\beta$
which characterize the stretch-luminosity and color-luminosity relations
in addition to the model parameters and the nuisance parameter $\mathcal{M}_B$ which
accounts for some combination of the absolute magnitude of a fiducial SN Ia and the Hubble constant.
Because the nuisance parameter $\mathcal{M}_B$
incorporates the absolute magnitude and the Hubble constant, and the normalization of the magnitude is arbitrary,
so we marginalize over it in the SNe Ia data fitting process, and $H_0$ is not a fitting parameter for the SNe Ia data.

The JLA sample combines the data from SNLS3 and three years of the SDSS-II SNe survey
with the joint light-curve analysis by the SDSS-II and SNLS collaborations, and
it includes a total of 740 spectroscopically confirmed SNe Ia with the improved photometric calibration \cite{jla}.
For the fitting to the JLA data, we also need to add two more nuisance parameters $\alpha$ and $\beta$.

The Union2.1 SNe Ia data consists of 580 SNe Ia which augments the Union2 compilation \cite{union2}
with 14 new SNe Ia from the HST Cluster Supernova Survey \cite{union2.1}
and 9 low redshift SNe Ia from the CSP \cite{Contreras:2009nt},
and it uses the SALT2 light-curve fitter.
There are 256 SNe Ia contained in both the SNLS3 and Union2.1 compilations.

For the BAO data, we use the measurements from the 6dFGS survey \citep{6dfgs}, the galaxy clustering
in the Baryon Oscillation Spectroscopic Survey (BOSS) \cite{Anderson:2012sa},
the distribution of galaxies in the SDSS survey \cite{wjp},
the WiggleZ dark energy survey \citep{wigglez}, and Ly$\alpha$ forest of high-redshift quasars in
BOSS survey \cite{ngbusca}.
Percival {\em et al} measured the distance ratio $d_{z}= r_{s}(z_{d})/D_{V}(z)$
at two redshifts $z=0.2$ and $z=0.35$ by fitting to the power spectra
of luminous red galaxies and main-sample galaxies in the SDSS \cite{wjp}.
Beutler {\em et al} derived that $d_{0.106}=0.336\pm 0.015$
from the 6dFGS measurements \cite{6dfgs}. The BOSS survey gave $d_{0.57}^{-1}=13.67\pm 0.22$ \cite{Anderson:2012sa}.
The WiggleZ dark energy survey measured the acoustic parameter
$A(z)=D_V(z)\sqrt{\Omega_m H_0^2}/z$
at three redshifts $z=0.44$, $z=0.6$ and $z=0.73$ \cite{wigglez}.
Busca {\em et al} reported the radial BAO data $\Delta z(z)=H(z)r_s(z_d)/c=0.11404\pm 0.00396$
at the redshift $z=2.3$ by detecting the BAO in
the Ly$\alpha$ forest of high-redshift quasars from the BOSS Survey \cite{ngbusca}.

For the Hubble parameter $H(z)$ data, we use the $H(z)$ data at 11 different
redshifts obtained from the differential ages of
passively evolving galaxies \cite{hz1,hz1a}, two Hubble parameter data $H(z=0.24)=79.69\pm 2.65$ km$\,$s$^{-1}$$\,$Mpc$^{-1}$
and $H(z=0.43)=86.45\pm 3.68$ km$\,$s$^{-1}$$\,$Mpc$^{-1}$
determined by taking the BAO scale as a standard ruler in the radial direction \citep{hz2}, and the $H(z)$ data
at eight different redshifts
obtained from the differential spectroscopic evolution of early-type
galaxies as a function of redshift \cite{hz3}. The total number of $H(z)$ data is 21.
Although the quality of the data is not good, but it helps constrain the behaviour of dark energy
because it depends on $w(z)$ by its first integral.

For the {\em Planck} data \cite{planck13}, we use the {\em Planck} temperature power spectrum data together
with the nine years of WMAP polarization low-multipole likelihood which was called {\em Planck}+WP in \cite{planck13},
hereafter we call this data {\em Planck} data for short.
For the SSLCPL model discussed in section 3, a flat prior over the range $[-3,\ 2]$ was assumed for the parameter $w_0$.
For the CPL model discussed in section 4, we choose the prior ranges  $[-3,\ 2]$ for $w_0$
and $[-10,\ 3]$ for $w_a$. Larger prior ranges were imposed because the
likelihood of $w_0$ has a long tail with $w_0>-0.3$ as can be seen from our results in section 4.

\section{SSLCPL parametrization}

\begin{figure}[htp]
\centerline{\includegraphics[width=0.9\textwidth]{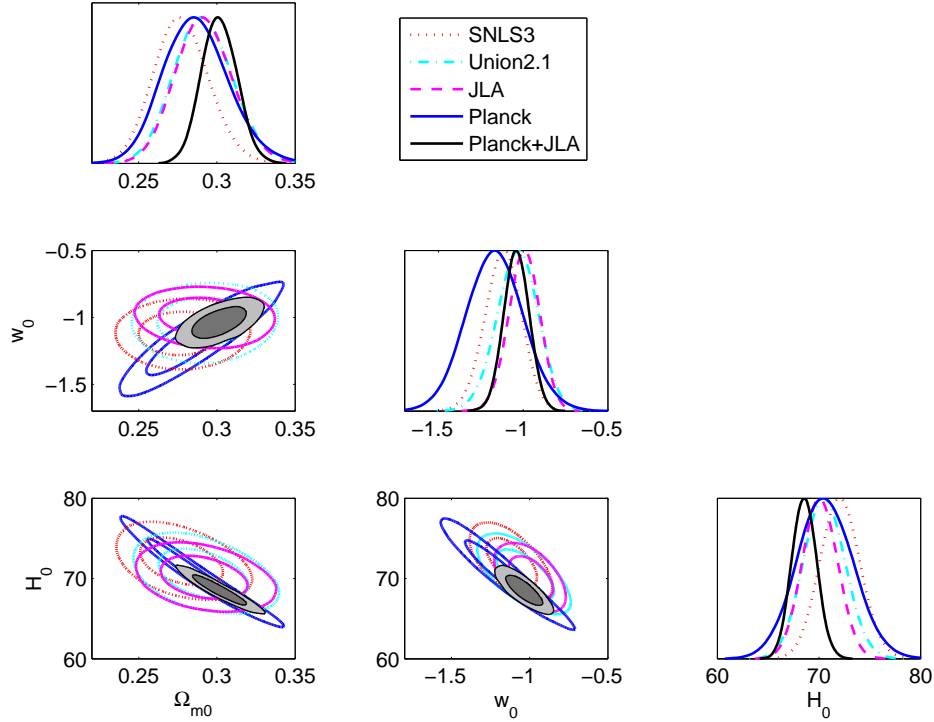}}
\caption{The marginalized $1\sigma$ and $2\sigma$ constraints on the flat SSLCPL model.
The red dotted line is for the combination SNLS3+BAO+$H(z)$, the dash-dotted cyan line is for
the combination Union2.1+BAO+$H(z)$, the dashed magenta line is for
the combination JLA+BAO+$H(z)$, the solid blue line is for the combination Planck+BAO+$H(z)$,
and the solid black line is the combination JLA+Union2.1+BAO+$H(z)$.}
\label{sslcpl_tri}
\end{figure}

To study the effect of dynamical dark energy model on the tensions among different data, we first consider
the SSLCPL parametrization \cite{gong1212,Gong:2013bn} which approximates the dynamics of
general thawing scalar fields \cite{Caldwell:2005tm}
over a large redshift range with only
one free parameter $w_0$,
and reduces to $\Lambda$CDM model when the parameter $w_0=-1$.
The model does not differ much from $\Lambda$CDM model, it approximates a wide class
of thawing scalar fields with the equation
of state $w(a)=w_0+w_a(1-a)$, and the parameter $w_a$ is a function of $w_0$ and $\Omega_{m0}$.
For the flat case, the equation of state of the SSLCPL model is \cite{gong1212,Gong:2013bn}
\begin{eqnarray}
\label{waeq1}
w(a)=w_0+6(1+w_0)\frac{(\Omega_{\phi 0}^{-1}-1)[\sqrt{\Omega_{\phi0}}-\tanh^{-1}(\sqrt{\Omega_{\phi0}})]}
{\Omega_{\phi 0}^{-1/2}-(\Omega_{\phi 0}^{-1}-1)\tanh^{-1}(\sqrt{\Omega_{\phi0}})}(1-a),
\end{eqnarray}
where $\Omega_{\phi 0}=1-\Omega_{m0}$.
By fitting the flat SSLCPL model to the SNLS3 SNe Ia data alone,
the Union2.1 SNe Ia data alone, the JLA SNe Ia data alone,
the BAO data alone and the $H(z)$ data alone, we find that
the values of $\Omega_{m0}$ are all consistent except that
the SNLS3 SNe Ia result is marginally consistent at the $1\sigma$ level.
The SNLS3 SNe Ia data alone prefers smaller values of $\Omega_{m0}$ and bigger values of $w_0$.
Next, we combine different SNe Ia data and {\em Planck} data with the BAO and $H(z)$ data and
the constraints on the model parameters from different combinations
are shown in table \ref{table1} and figure \ref{sslcpl_tri}.
By combining the BAO and $H(z)$ data with different SNe Ia and {\em Planck} data,
we find that the constraints on the parameters are all consistent as seen from table \ref{table1} and figure \ref{sslcpl_tri}.
In particular, the value of $H_0$ from different data agrees at the $1\sigma$ level for the SSLCPL model.
The combination of Planck+BAO+$H(z)$ prefers
more negative value of $w_0$ which is marginally consistent with $\Lambda$CDM model at the $1\sigma$ level,
and the error bar on $H_0$ is doubled comparing with the constraint on $\Lambda$CDM model.
The combination of
SNLS3+BAO+$H(z)$ gives a lower value of $\Omega_{m0}$ and a higher value of $H_0$
and the result is in tension with $\Lambda$CDM model. With the improved photometric calibration,
the combination of JLA+BAO+$H(z)$ not only tightens the constraint on $w_0$
a little bit but also eliminates the tension with $\Lambda$CDM model.
Both the combinations of Union2.1+BAO+$H(z)$ and JLA+BAO+$H(z)$ are consistent with $\Lambda$CDM model.
Furthermore, the constraint on $H_0$ from JLA+BAO+$H(z)$ is consistent with both the local
measurements and the {\em Planck} result because of the larger error bar on $H_0$.

\begin{table}[pht]
\caption{The marginalized $1\sigma$ constraints on the flat SSLCPL model.}
\begin{tabular}{ccccc}
\hline
Data & $\Omega_{m0}$ & $w_0$ & $H_0$  \\ \hline
Union2.1+BAO+$H(z)$ & $0.291_{-0.019}^{+0.018}$  & $-1.03_{-0.12}^{+0.11}$ & $70.5_{-1.9}^{+2.0}$  \\\hline
SNLS3+BAO+$H(z)$ & $0.277_{-0.017}^{+0.018}$  & $-1.12\pm 0.11$ & $72.1_{-2.0}^{+1.9}$  \\\hline
JLA+BAO+$H(z)$ & $0.292_{-0.019}^{+0.017}$  & $-1.0\pm 0.09$ & $70.1\pm 1.7$  \\\hline
Planck+BAO+$H(z)$ & $0.287^{+0.019}_{-0.023}$ & $-1.17^{+0.17}_{-0.16}$ & $70.6^{+2.7}_{-2.6}$ \\\hline\hline
Union2.1+Planck+BAO+$H(z)$ & $0.292^{+0.013}_{-0.015}$ & $-1.12^{+0.10}_{-0.11}$ & $69.9^{+1.7}_{-1.8}$ \\\hline
SNLS3+Planck+BAO+$H(z)$ & $0.287^{+0.012}_{-0.013}$ & $-1.15\pm 0.09$ & $70.4^{+1.5}_{-1.4}$ \\\hline
JLA+Planck+BAO+$H(z)$ & $0.302^{+0.011}_{-0.012}$ & $-1.04\pm 0.07$ & $68.6^{+1.1}_{-1.2}$ \\\hline
\end{tabular}
\label{table1}
\end{table}

Since both SNLS3+BAO+$H(z)$ and Planck+BAO+$H(z)$ prefers lower values of $w_0$, when
we use the combination of SNLS3+Planck+BAO+$H(z)$,
we expect that a tension with $\Lambda$CDM model remains,
in fact the result is $w_0=-1.15\pm 0.09$ which differs from $\Lambda$CDM model at more than $1\sigma$ level.
In addition to this tension, the value of
$H_0$ is also in tension with the {\em Planck} result at about $2\sigma$ level. Replacing
the SNLS3 data with the improved JLA data, we get $\Omega_{m0}=0.302^{+0.011}_{-0.012}$,
$w_0=-1.04\pm 0.07$ and $H_0=68.6^{+1.1}_{-1.2}$ with $\chi^2=10518.15$.
The constraints on the flat $\Lambda$CDM model from the
combined JLA+Planck+BAO+$H(z)$ is $\Omega_{m0}=0.305\pm 0.09$
and $H_0=68.0^{+0.7}_{-0.6}$ with $\chi^2=10518.02$.
The results suggest that the
systematics in SNLS3 data is the main reason for the tensions.
However, the value of $H_0=68.6^{+1.1}_{-1.2}$ km$\,$s$^{-1}$$\,$Mpc$^{-1}$
is now in tension with the local measurements at about $1.9\sigma$ level although the tension is partially alleviated.
Therefore, when dynamical dark energy model is used, the tension on $H_0$ may get partially relieved due to
larger error bar on $H_0$ with the inclusion of more fitting parameters, but $\Lambda$CDM model seems not to be the
reason of the tension.
By using the combination of Union2.1+Planck+BAO+$H(z)$, we get $w_0=-1.12^{+0.10}_{-0.11}$ which is in tension
with $\Lambda$CDM model at the $1\sigma$ level.

\section{CPL parametrization}

Fitting the flat CPL model to the SNLS3 SNe Ia
data, we get $\Omega_{m0}=0.32\pm 0.08$, $w_0=-0.7\pm 0.3$
and $w_a=-4.9^{+4.2}_{-4.5}$ which is marginally consistent with $\Lambda$CMD model at the $1\sigma$ level.
Fitting the flat CPL model to the combined Planck+BAO data,
we get $\Omega_{m0}=0.307^{+0.03}_{-0.042}$, $w_0=-0.90_{-0.49}^{+0.37}$, $w_a=-0.76^{+1.62}_{-0.94}$  and $H_0=68.9^{+3.9}_{-4.0}$ km$\,$s$^{-1}$$\,$Mpc$^{-1}$
which is consistent with the flat $\Lambda$CDM model and those from SNLS3 SNe Ia data alone,
the result is also consistent with that in \cite{Xia:2013dea}.
The combined BAO+Planck data also gives much tighter constraint
on the variation of dark energy parameter $w_a$ than SNLS3 SNe Ia data due to the
tighter constraint on $\Omega_{m0}$, and it prefers a more negative value of $w_0$.
The BAO and {\em Planck} data do not prefer dark energy models with rapid change of $w(z)$,
so they constrain $w_a$ to be a smaller range around zero.

\begin{figure}[htp]
\centerline{\includegraphics[width=0.9\textwidth]{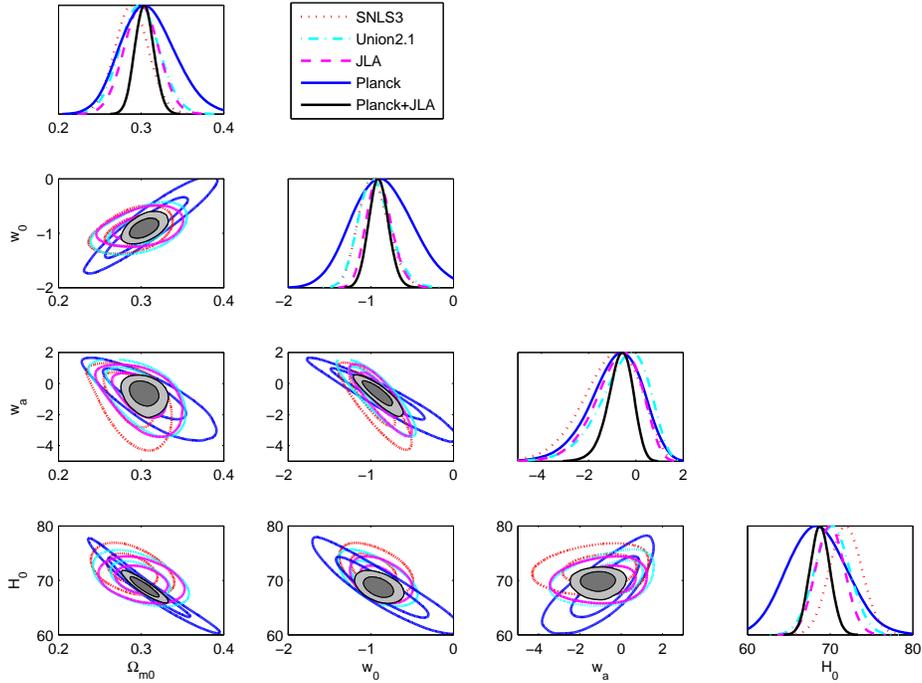}}
\caption{The marginalized $1\sigma$ and $2\sigma$ constraints on the flat CPL model.
The red dotted line is for the combination SNLS3+BAO+$H(z)$, the dash-dotted cyan line is for
the combination Union2.1+BAO+$H(z)$, the dashed magenta line is for
the combination JLA+BAO+$H(z)$, the solid blue line is for the combination Planck+BAO+$H(z)$,
and the solid black line is the combination JLA+Union2.1+BAO+$H(z)$.}
\label{cpl_tri}
\end{figure}

Fitting the flat CPL model to the combined SNLS3+BAO+$H(z)$ data,
we get $\Omega_{m0}=0.289 \pm 0.020$, $w_0=-0.96\pm 0.18$, $w_a=-1.13\pm 1.20$ and $H_0=71.9^{+2.0}_{-1.9}$ km$\,$s$^{-1}$$\,$Mpc$^{-1}$
which differs by about $2\sigma$ from the {\em Planck} result.
Fitting the flat CPL model to the combined Union2.1+BAO+$H(z)$ data,
we get $\Omega_{m0}=0.298^{+0.023}_{-0.022}$, $w_0=-0.95^{+0.17}_{-0.21}$,
$w_a=-0.58^{+1.27}_{-0.75}$ and $H_0=70.3\pm 2.0$ which are consistent with
both the $\Lambda$CDM model and the {\em Planck} result.
The constraints from Union2.1 data are also consistent with SNLS3 data.
Replacing the SNLS3 data by the improved JLA data, we get
$\Omega_{m0}=0.30\pm 0.02$, $w_0=-0.89^{+0.13}_{-0.17}$, $w_a=-0.82^{+1.16}_{-0.75}$ and $H_0=70.0\pm 1.7$
which is consistent with both SNLS3 and Union2.1 data,
and the results are also consistent with $\Lambda$CDM model. The JLA data
prefers larger value of $w_0$. Comparing with the constraints on the SSLCPL
model, we find that the constraints on $H_0$ are almost the same even though we have one more parameter $w_a$.
Fitting the CPL model to the combined Placnk+BAO+$H(z)$ data,
we get $\Omega_{m0}=0.309^{+0.029}_{-0.036}$, $w_0=-0.86^{+0.34}_{-0.39}$, $w_a=-0.84^{+1.23}_{-0.90}$ and $H_0=68.6^{+3.5}_{-3.6}$ km$\,$s$^{-1}$$\,$Mpc$^{-1}$
which is consistent with the $\Lambda$CDM model.
The results are summarized in table \ref{tablecpl} and shown in figure \ref{cpl_tri}.
With one more parameter on $w(z)$, the Placnk+BAO+$H(z)$ data prefers larger value of $\Omega_{m0}$
and smaller value of $H_0$, and the constraints on the cosmological parameters become a little looser.
Although the constraints from different SNe Ia data and {\em Planck} data are consistent,
the JLA data is more compatible with the {\em Planck} data than the Union2.1 data,
and the JLA data give much tighter constraints than the {\em Planck} data as seen from table \ref{tablecpl} and figure \ref{cpl_tri}.

\begin{table}[pht]
\caption{The marginalized $1\sigma$ constraints on CPL model.}
\begin{tabular}{ccccc}
\hline
Data & $\Omega_{m0}$ & $w_0$ & $w_a$ & $H_0$\\ \hline
Union2.1+BAO+$H(z)$ & $0.298^{+0.023}_{-0.022}$  & $-0.95^{+0.17}_{-0.21}$ & $-0.58_{-0.75}^{+1.27}$ & $70.3\pm 2.0$   \\\hline
SNLS3+BAO+$H(z)$ & $0.289\pm 0.020$  & $-0.96\pm 0.18$ & $-1.13\pm 1.20$ & $71.9^{+2.0}_{-1.9}$   \\\hline
JLA+BAO+$H(z)$ & $0.30\pm 0.02$  & $-0.89^{+0.13}_{-0.17}$ & $-0.82_{-0.75}^{+1.16}$ & $70.0\pm 1.7$  \\\hline
Planck+BAO+$H(z)$ & $0.309^{+0.029}_{-0.036}$ & $-0.86^{+0.34}_{-0.39}$ & $-0.84^{+1.23}_{-0.90}$ & $68.6^{+3.5}_{-3.6}$  \\\hline\hline
Union2.1+Planck+BAO+$H(z)$ & $0.298^{+0.015}_{-0.017}$ & $-0.98^{+0.17}_{-0.2}$ & $-0.49^{+0.79}_{-0.54}$ & $69.5^{+1.8}_{-1.7}$ \\\hline
SNLS3+Planck+BAO+$H(z)$ & $0.29^{+0.012}_{-0.013}$ & $-1.02^{+0.13}_{-0.14}$ & $-0.51^{+0.69}_{-0.48}$ & $70.5\pm 1.4$  \\\hline
JLA+Planck+BAO+$H(z)$ & $0.304\pm 0.011$ & $-0.90^{+0.11}_{-0.12}$ & $-0.68^{+0.60}_{-0.44}$ & $68.8^{+1.1}_{-1.2}$  \\\hline
\end{tabular}
\label{tablecpl}
\end{table}

Fitting the flat CPL model to the combined Union2.1+Planck+BAO+$H(z)$ data, we get $\Omega_{m0}=0.298^{+0.015}_{-0.017}$,
$w_0=-0.98^{+0.17}_{-0.2}$, $w_a=-0.49^{+0.79}_{-0.54}$,
and $H_0=69.5^{+1.8}_{-1.7}$ km$\,$s$^{-1}$$\,$Mpc$^{-1}$ which is consistent with $\Lambda$CDM model.
Fitting the flat CPL model to the combined SNLS3+Planck+BAO+$H(z)$ data, we get $\Omega_{m0}=0.290^{+0.012}_{-0.013}$,
$w_0=-1.02^{+0.13}_{-0.14}$, $w_a=-0.51^{+0.69}_{-0.48}$,
and $H_0=70.5\pm 1.4$ km$\,$s$^{-1}$$\,$Mpc$^{-1}$ which differs by about $1.7\sigma$ from the {\em Planck} value for the flat $\Lambda$CDM model.
Fitting the flat CPL model to the combined JLA+Planck+BAO+$H(z)$ data, we get $\Omega_{m0}=0.304\pm 0.011$,
$w_0=-0.90^{+0.11}_{-0.12}$, $w_a=-0.68^{+0.60}_{-0.44}$,
and $H_0=68.8^{+1.1}_{-1.2}$ km$\,$s$^{-1}$$\,$Mpc$^{-1}$ with $\chi^2=10516.08$ which is consistent with the results obtained
in \cite{jla} with distance prior from {\em Planck} data.
For the Union2.1 data, the results are consistent with $\Lambda$CDM model due to larger error bars
on the constraints. Since the JLA data is more compatible with the {\em Planck} data, the combined JLA+Planck+BAO+$H(z)$ give
higher value of $\Omega_{m0}$ and lower value of $H_0$ and the error bars on $\Omega_{m0}$ and $H_0$ are small, the value of $H_0$ is
consistent with the {\em Planck} result, but it is in tension with the local measurements at about $1.9\sigma$ level.
Furthermore, the $\Lambda$CDM model with $w_0=-1$ and $w_a=0$ is only marginally consistent with the results
from the combined JLA+Planck+BAO+$H(z)$ data at the $1\sigma$ level.

\section{Conclusions}

The {\em Planck} first year result further supports the concordance $\Lambda$CDM model with
$\Omega_{m0}=0.315\pm 0.017$ and $H_0=67.3\pm 1.2$ km$\,$s$^{-1}$$\,$Mpc$^{-1}$. Although the tension on $\Omega_{m0}$
is due to the systematics in SNLS3 SNe Ia data, the tension on $H_0$ with the local measurements still exists.
We study the reason behind it by lifting the restriction on the dark energy model.
In particular, we extend the flat $\Lambda$CDM model to flat SSLCPL model and flat CPL model.
For the flat SSLCPL model, we find that {\em Planck} data is consistent with BAO and $H(z)$ data,
and the constraints $\Omega_{m0}=0.287^{+0.019}_{-0.023}$ and $H_0=70.6^{+2.7}_{-2.6}$ km$\,$s$^{-1}$$\,$Mpc$^{-1}$
from the combined Planck+BAO+$H(z)$ data are consistent with the local measurements and
the {\em Planck} results for the concordance $\Lambda$CDM model, but the value
of $w_0=-1.17^{+0.17}_{-0.16}$ is marginally consistent with $w_0=-1$ at the $1\sigma$ level.
Both the Union2.1 and JLA data are consistent with $\Lambda$CDM model and the {\em Planck} data,
the SNe Ia data give much better constraints on $w_0$ than the {\em Planck} data.
Due to larger error bars on $H_0$, the constraints
on $H_0$ from Union2.1+BAO+$H(z)$, JLA+BAO+$H(z)$ and Planck+BAO+$H(z)$ data are
consistent with both the local measurements and the {\em Planck} result.
Combing the Union2.1 SNe Ia data with the {\em Planck} data, we get
$\Omega_{m0}=0.292^{+0.013}_{-0.015}$, $w_0=-1.12^{+0.10}_{-0.11}$ and $H_0=69.9^{+1.7}_{-1.8}$.
The value of $w_0$ is in tension with the $\Lambda$CDM value $w_0=-1$ at the $1\sigma$ level
although the value of $H_0$ is consistent with both {\em Planck}
result and the local measurements due to larger error bars.
For the combined JLA+Planck+BAO+$H(z)$ data, there is no tension with the $\Lambda$CDM model, but
the value of $H_0=68.6^{+1.1}_{-1.2}$ is in tension with the local measurements by about $1.9\sigma$. The result suggests
that with reduced error bar on the cosmological parameters due to higher quality of data, the
tension on $H_0$ remains.

The constraints on SSLCPL model from the combined Planck+BAO+$H(z)$ data
are marginally consistent with $\Lambda$CDM model although
the SSLCPL is a simple extension of $\Lambda$CDM model.
Therefore, we further ease the restriction on the dark energy model by considering the flat CPL parametrization.
The results show that no tension exists among different data for the flat CPL model and the results from different
data in combinations with BAO+$H(z)$ data
are consistent with $\Lambda$CDM model.
The combined JLA+BAO+$H(z)$ data give much tighter constraints on $\Omega_{m0}$,
$w_0$ and $H_0$ than the combined Planck+BAO+$H(z)$ data. Because we have more parameters,
the error bars on the parameters are usually larger,
that may be the reason for the consistency of the results.
The error bars can be further reduced when we combine the SNe Ia data with the {\em Planck} data
because they have different degeneracy directions.
By fitting the flat CPL model to the combined Union2.1+Planck+BAO+$H(z)$ data,
we get $\Omega_{m0}=0.298^{+0.015}_{-0.017}$,
$w_0=-0.98^{+0.17}_{-0.2}$, $w_a=-0.49^{+0.79}_{-0.54}$,
and $H_0=69.5^{+1.8}_{-1.7}$ km$\,$s$^{-1}$$\,$Mpc$^{-1}$ which is consistent with $\Lambda$CDM model,
but the error bar is still large.
Fitting the flat CPL model to the combined JLA+Planck+BAO+$H(z)$ data, we get $\Omega_{m0}=0.304\pm 0.011$,
$w_0=-0.90^{+0.11}_{-0.12}$, $w_a=-0.68^{+0.60}_{-0.44}$,
and $H_0=68.8^{+1.1}_{-1.2}$ km$\,$s$^{-1}$$\,$Mpc$^{-1}$. With the addition of {\em Planck} data to the JLA data,
we get $w_a<0$ at the $1\sigma$ level,
the constraints on $\Omega_{m0}$, $w_a$ and $H_0$ was improved about 40\%, the result
is only marginally consistent with $\Lambda$CDM model and $H_0$ is in tension with the local measurements.
Note that when the distance priors from the {\em Planck} data were used,
it was found that $w_a=-0.336\pm 0.552$
which is consistent with $\Lambda$CDM value $w_a=0$ \cite{jla}.
Because the constraints on $H_0$ are almost unchanged for different SNe Ia
data when we use CPL model in place of SSLCPL model,
so the tension on $H_0$ is not caused by the fitting model.
Fitting the combined JLA+Planck+BAO+$H(z)$ data to the
flat $\Lambda$CDM model, we get $\Omega_{m0}=0.305\pm 0.09$
and $H_0=68.0^{+0.7}_{-0.6}$ which differs about $2.3\sigma$ from the local measurements.
In terms of Akaike information criteria (AIC) \cite{aic}, $\Lambda$CDM model is favored.

From the above analysis, we
see that the constraints on $H_0$ from SSLCPL and CPL models are almost the same, and
the tension on $H_0$ with the local measurements exists
when the error bar on $H_0$ is tightened to be around 1. The tension on $H_0$ seems
not come from dark energy models, it seems that both the local measurements and the {\em Planck}
data have some unknown systematics. For example, by using a map-based foreground cleaning procedure
based on a combination of 353 GHz and 545 GHz maps, it was found that $H_0=68.1\pm 1.1$ km$\,$s$^{-1}$$\,$Mpc$^{-1}$
for the {\em Planck} data \cite{Spergel:2013rxa}.
By revising the geometric maser distance to NGC 4258,
the Hubble constant was lowered to be $H_0=70.6\pm 3.3$ km$\,$s$^{-1}$$\,$Mpc$^{-1}$ \cite{Efstathiou:2013via}.
So the tension may disappear if systematics were properly accounted for.

In conclusion, there is no tension among different data. With
the improved photometric calibration for the SNLS3 SNe Ia data, the JLA data give
tighter constraints on the cosmological parameters.
The combined JLA+BAO+$H(z)$ data constrain $w_0$ much better than the combined Planck+BAO+$H(z)$ data,
and the combined JLA+Planck+BAO+$H(z)$ data constrains $H_0$ to be better than 2\%.
Although dynamical dark energy model with $w_a<0$ is preferred at the $1\sigma$ level
by the combined JLA+Planck+BAO+$H(z)$ data, $\Lambda$CDM model is still favored in term of AIC.
The tension on $H_0$ with the local measurements remains and it is not caused by the fitting model.
Further studies on the systematics of both {\em Planck} data and the local measurements on $H_0$ are needed.

\begin{acknowledgments}
This work was partially supported by
the National Basic Science Program (Project 973) of China under
grant No. 2010CB833004, the NNSF of China under grant nos. 10935013 and 11175270,
the Program for New Century Excellent Talents in University under grant no. NCET-12-0205
and the Fundamental Research Funds for the Central Universities under grant no. 2013YQ055.
\end{acknowledgments}


\providecommand{\newblock}{}

\end{document}